\documentclass[preprint]{aa}  

\usepackage{graphicx}
\usepackage{txfonts}
\usepackage{natbib}
\bibpunct{(}{)}{;}{a}{}{,} 

\def\amin{\ifmmode^{\prime}\else$^{\prime}$\fi} 
\def\asec{\ifmmode ^{\prime\prime}\else$^{\prime\prime}$\fi} 
\def\asec{\ifmmode ^{\prime\prime}\else$^{\prime\prime}$\fi}

\def\it{\sl}
\def\degs{\ifmmode ^{\circ}\else$^{\circ}$\fi}
\def\amin{\ifmmode ^{\prime}\else$^{\prime}$\fi}
\def\asec{\ifmmode ^{\prime\prime}\else$^{\prime\prime}$\fi}
\def\fm{\hbox{$.\!\!^{\rm m}$}}            
\def\farcs{\hbox{$.\!\!^{\prime\prime}$}}  
\def\famin{\hbox{$.\!\!^{\prime}$}} 

\def\psr{PSR~J1124-5916\ }

\def\degs{\ifmmode ^{\circ}\else$^{\circ}$\fi}
\def\amin{\ifmmode ^{\prime}\else$^{\prime}$\fi}

\def\eqalign#1{\null\,\vcenter{\openup1\jot \m@th
   \ialign{\strut\hfil$\displaystyle{##}$&$\displaystyle{{}##}$\hfil
   \crcr#1\crcr}}\,}
\begin{document}
   \title{ 
The PSR J1124-5916 wind nebula in the
near-infrared
   \thanks{Based 
on observations made with ESO telescope at the Paranal Observatory under
Programme 084.D-0617(A).}}

   \author{Sergey V.~Zharikov\inst{1}, Dmitry A.
   ~Zyuzin\inst{2}, Yuri A.~Shibanov\inst{2,3},
   Ronald E.~Mennickent\inst{4}}

   \institute{Instituto de Astronom\'{i}a, Universidad Nacional Aut\'{o}noma de M\'{e}xico, Ensenada, Mexico\\
              \email{zhar@astrosen.unam.mx}
             \and
             Ioffe Physical Technical Institute, Politekhnicheskaya 26, St. Petersburg, 194021, Russia\\
             \email{dima\_zyuz@mail.ru}    
            \and 
	    St. Petersburg State Polytechnical Univ., Politekhnicheskaya 29, St. Petersburg, 195251, Russia\\      
	     \email{shib@astro.ioffe.ru}
	     \and 
             Departamento de Astronom\'{i}a, Universidad de Concepci\'{o}n, Casilla 160-C, Concepci\'{o}n, Chile \\  
             \email{rmennick@astro-udec.cl}           
             }

   \date{Received ...; accepted ...}

   \abstract
   {The young radio pulsar  J1124-5916 is associated with
   a Cas A like supernova remnant G292.0+1.8. It powers a compact torus-like pulsar wind 
nebula with a jet  first detected in X-rays and then  
identified in the optical and mid-infrared.}
   {We  carried out  deep    near-infrared 
observations  of the pulsar field 
to identify the pulsar and its 
nebula in this range. }
    {
   The direct imaging mode of the NACO adaptive optics instrument
at the ESO VLT  in the $H$ and $K_s$ bands was used. }
    {In both   bands we detected  a  faint,
    $H=21.30(10)$ and $K_s=20.45(10)$, extended elliptical 
object, whose  center position is  
consistent  with the X-ray position of the pulsar. 
The morphology of the object and the orientation of its major axis are in 
a good agreement with those observed for the pulsar
torus-like nebula in
the mid-infrared, optical, and X-rays. This suggests
that  it is the near-infrared counterpart  of the
nebula. The measured fluxes compiled with the data
in other ranges show a  complicated unabsorbed power law
spectrum of the torus-like nebula with several steep 
breaks between the near-infrared and mid-infrared,
the optical and X-rays, and possibly in the
mid-infrared.  This implies a multiple   relativistic
particle population 
responsible for the
synchrotron emission of the nebula in
different spectral ranges.  
We have not resolved the pulsar counterpart from its
nebula and  
place only upper limits on its brightness,  
$H\geq 23.9$ and $K_s \geq 22.7$.
Based on that,  its contribution to the total
near-infrared flux of the pulsar+nebula system is $\le
10\%$, which is comparable with the expected
contribution in the optical. }
{}


\authorrunning{S. Zharikov,  et al.}
\titlerunning{ The PSR J1124-5916 wind nebula in the
near-infrared}
   \maketitle

%
\section{Introduction}
\label{int}
 \psr was discovered  in the
  radio  by \citet{2002ApJ...567L..71C} and in X-rays
 by \citet{2001ApJ...559L.153H, 2003ApJ...591L.139H} and then  detected  
 in $\gamma$-rays by \citet{2010ApJS..187..460A}. 
It is associated  with  the supernova 
  remnant (SNR)  G292.0+1.8 (MSH 11-54).  
 The characteristic age of
the pulsar $\tau$ $\approx$2900~yr and  its spin-down luminosity
  $\dot{E}$ $\approx$1.2$\times10^{37}$ ergs~s$^{-1}$  
rank this pulsar as the sixth youngest and the  twelfth most energetic
among all rotation-powered pulsars  known. 
The pulsar age  is consistent with 2700--3700~yr age of  G292.0+1.8 \citep{2002ApJ...567L..71C,2005ApJ...619..839C}.  
The  remnant low distance limit  of 3.2 kpc is based on the HI absorption to beyond the tangent point in 
this direction \citep{1975A&A....45..239C}. An upper  limit of 11 kpc follows 
from the pulsar dispersion measure ($DM$)  of 330 pc cm$^2$ and
\citet{1993ApJ...411..674T} model,  while  
$6.3\pm0.9$ kpc is considered as the most
plausible distance estimate \citep{2003ApJ...594..326G}.

As the Crab,  \psr~powers a compact, $\sim$5\asec, and
relatively bright torus-like X-ray
pulsar wind nebula (PWN) with a 
jet \citep{2001ApJ...559L.153H,2002ASPC..262..315S,2003ApJ...591L.139H,2007ApJ...670L.121P},   
which is surrounded by a fainter and more extended, 1\amin$\times$2\amin,  plerion
visible in the
radio  and X-rays \citep{2003ApJ...594..326G}.  
The torus-like PWN appears to be seen nearly edge-on.   
 Its brightest inner part  containing the
 pulsar was identified in the optical $VRI$  bands by
 \citet{2008A&A...492..805Z} and then in the mid-infrared (mid-IR)
at   4.5 and 8 $\mu$m \citep{2009A&A...508..855Z}.   The point-like pulsar was not
resolved  from the PWN in these ranges  and its contribution to the total pulsar+PWN system 
emission  was estimated to be $\la$ 20\%.  
The measured fluxes implied a double-knee spectral  break  in  the 
power-law spectrum of the system  between the
optical and X-rays. This is  distinct from the
Crab PWN whose spectrum shows only a single break in the same range.  
To  obtain more stringent constraints on
the optical-IR spectral energy distribution of the system,  
near-infrared (near-IR) observations  are necessary.  

Here we present the results  of such observations obtained 
in the $HK_s$ bands with the adaptive optic (AO) system at the ESO
Very Large Telescope (VLT).   We use also  archival  optical  and   mid-IR  data
obtained with the HST and  AKARI.      
The  observations and archival   data are 
described in Sect. \ref{obs}, the results are presented and discussed in
Sect.~\ref{res} \& \ref{dis} 
and concluded in
Sect. 
\ref{conc}. 

\section{VLT observations and  archival data}
\label{obs}
\begin{figure*}[t]
\setlength{\unitlength}{1mm}
\resizebox{12.cm}{!}{
\begin{picture}(120,71)(0,0)
\put (0,0) {\includegraphics[width=9.1cm, bb = 145 260 460 520, clip]{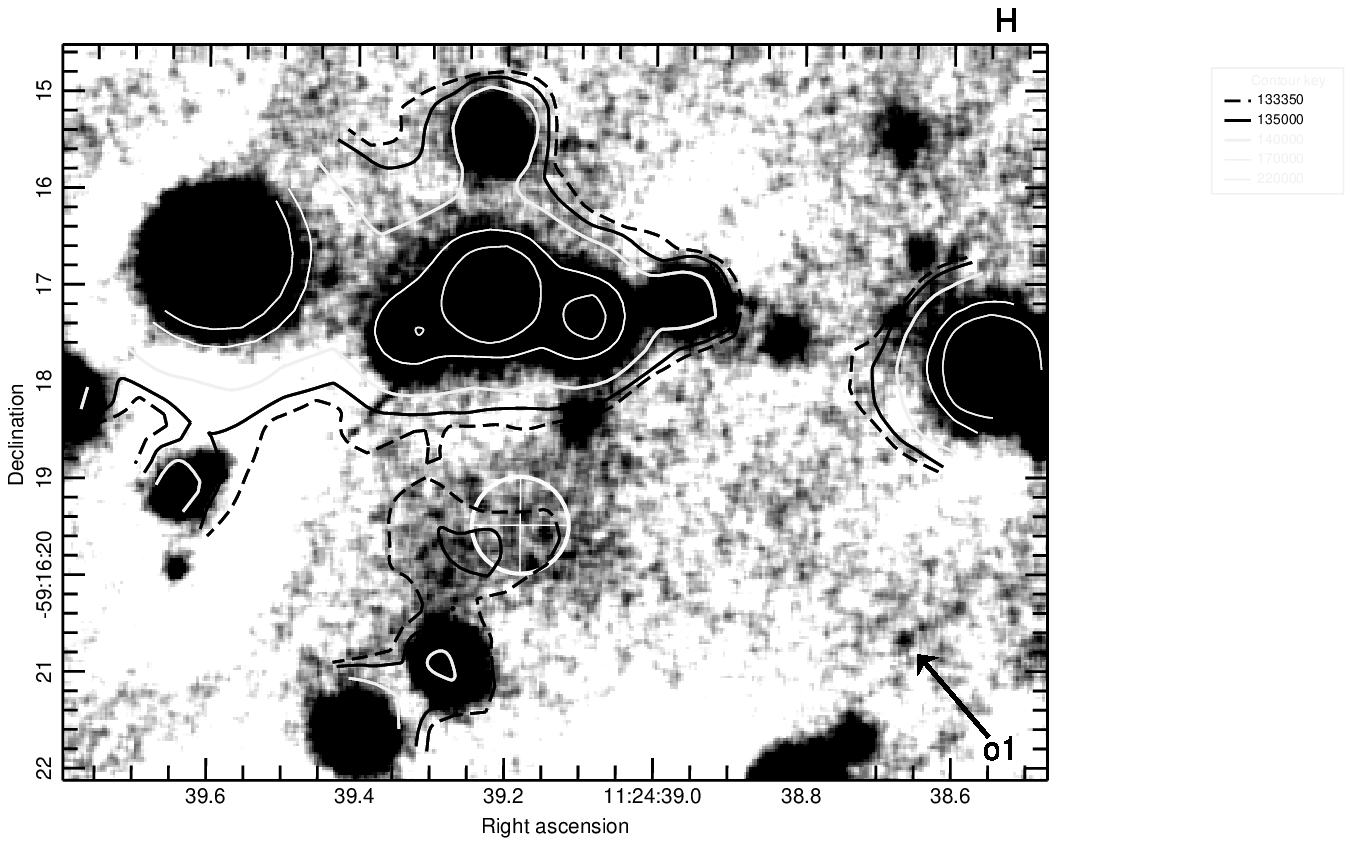}}
\put (90,0) {\includegraphics[width=9.1 cm, bb = 145 260 460 520, clip]{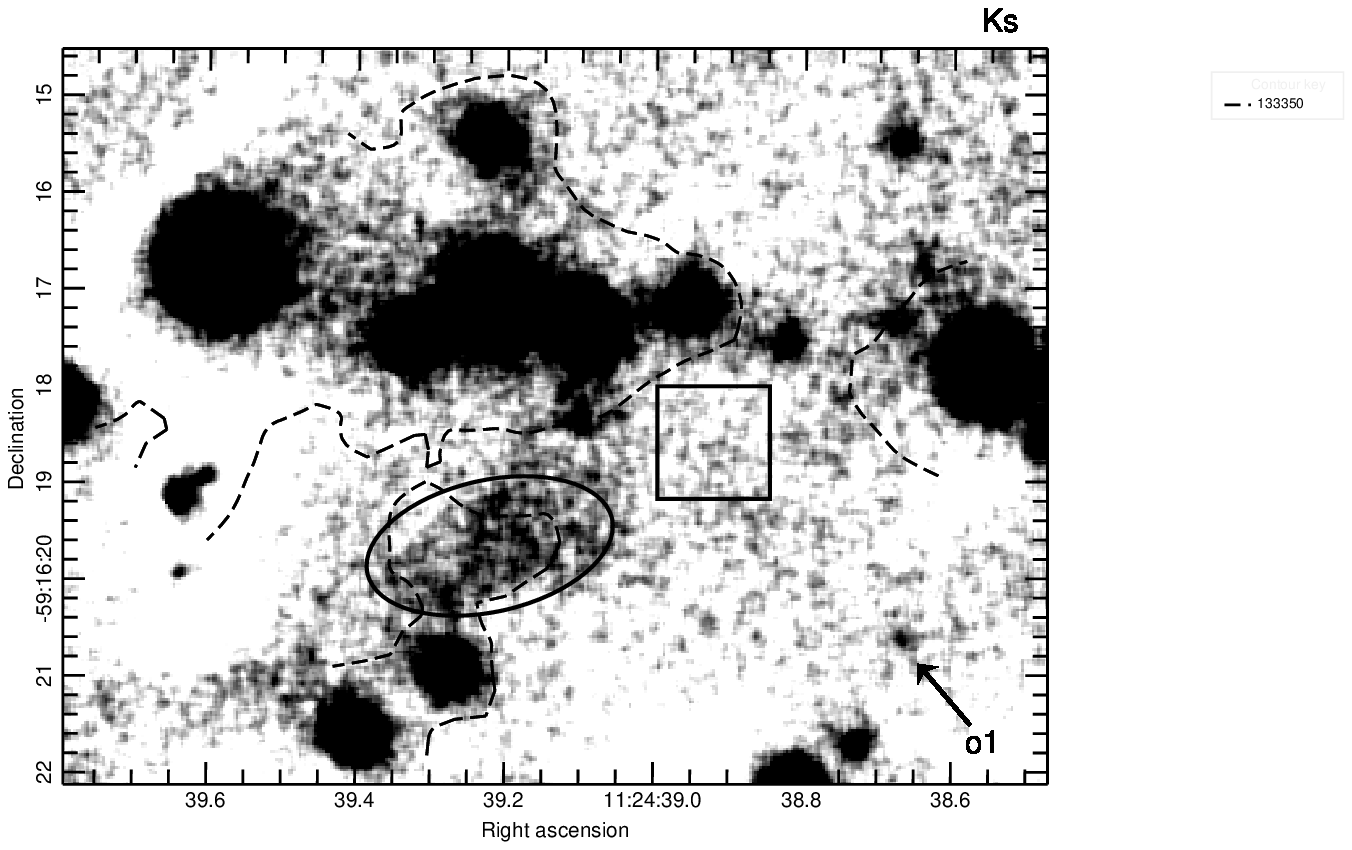}}
\end{picture}}
  \caption{VLT/NACO $H$ ({\sl left}) and $K_s$ ({\sl
  right}) images of the \psr~field smoothed with a Gaussian kernel of $3\times3$ pixels. 
  The cross and circle  in the $H$ image are  the X-ray pulsar position and    its 1$\sigma$  uncertainty. 
  The faint objects "o1"  is detected in both bands and  its
  magnitudes are accepted as   stellar-like object detection limits of the images.      
  The contours are overlaid from  the  $V$ band image  \citep{2008A&A...492..805Z}.  
  The ellipse and  rectangle in the $K_s$ image show   the aperture and background region, respectively, used
  for the PWN  photometry.    }
 \label{fig2}
\end{figure*}
\subsection{The VLT observations, data reduction, and
calibration}
\begin{table}[t]
\caption{Log of the VLT/NACO observations of PSR J1124-5916. }
\begin{tabular}{cllllll}
\hline\hline
 Date              &  Band         &  AM         &  Seeing  &NACO   & Zero-       \\
    (UT)             &                  &     &           & {\small FWHM}    & point      \\
   \hline \hline                 
    2010-03-20     &  $K_s$      &  1.23         &    0.84(5)& 0.15(3)        &  23.15(5)           \\
    2010-03-21     &  $K_s$      &  1.22         &    1.26(6)&  0.28(3)       &  23.12(3)           \\
         -"-                &  $K_s$      &  1.26         &    1.31(13)&  0.27(8)     &   -"-                       \\                               
   2010-03-23      &  $K_s$      &  1.27         &    0.88(10)&  0.38(6)     &  23.17(2)            \\
   2010-04-11      &  $K_s$      &  1.28         &    1.39(40)&  0.28(17)     &  23.17(2)            \\
   2010-06-18      &  $K_s$      &  1.36         &    0.67(6) &   0.34(4)     &  23.17(2)            \\ \hline
   2010-03-23      &  $H$    &  1.23         &    0.93(7) &0.40(3)        &  24.12(3)            \\
   2010-04-02      &  $H$    &  1.27         &  0.90(16) &0.24(6)         &  24.16(2)            \\
   2010-06-16      &  $H$    &  1.26         &  1.10(12) &0.42(5)         &  24.16(2)            \\
   2010-06-17      &  $H$    &  1.33         &  1.05(15) &0.49(3)         &                          \\
        -"-                 &  $H$   &  1.50         & 0.89(5)     &0.43(3)           &                             \\   
   2010-06-18      &  $H$    &  1.27         & 0.57(7)    &0.32(4)          &                             \\ 
   \hline \\
\end{tabular}\\
\begin{tabular}{l}
\begin{minipage}[t]{0.98\columnwidth}{\footnotesize
Each  line in the table corresponds to an observational  block  of ten
individual dithered exposures  with   integration  times  of  $114\times2$s.
Block averaged airmasses (AM), seeing and  AO  
 corrected FWHMs  (in  arcseconds),  and zero-points 
(in magnitude)  are shown; numbers  in brackets are 1$\sigma$
uncertainties.}
\end{minipage}
\end{tabular}
\label{t:log}
\end{table}
We observed PSR J1124-5916 in  service mode in  March
19, 20, 22, April 1, 11 and June  17, 18  2010 from the ESO
Paranal Observatory with the NAos COnica (NACO), 
the AO near-IR imager and spectrometer mounted at the VLT Yepun unit. In order to provide the best 
combination between angular resolution and sensitivity, we used the S27 camera which has a pixel 
scale of 0\farcs027 and a field of view (FOV) of
28\asec$\times$28\asec. As a reference for the AO correction we used 
the GSC-2 star S11131211160 ($V, R \sim16$ mag), located 
7\farcs9 away from our target. 
The visual dichroic element and wavefront sensor were used.
The observational log is given in Table~\ref{t:log}.  The
conditions were photometric with seeing  varying from
0\farcs6 to 1\farcs3.

 Day (darks, lamp flat-fields) and night  time (twilight
 flat-fields) calibration frames  
were taken daily as a part of the NACO calibration plan and
used to create "master" dark and flat-field frames for each
of the observational blocks (OBs). Standard data reduction of
science frames, including dark subtraction, flat-fielding, cosmic-ray
removal, and exposure  dithering correction, was performed. 
To remove   image backgrounds,    "super-flat" frames were
 created for each of the OBs  from a median combination    
 of   respective science images  where  stars were replaced by 
surrounding  background levels. The frames were normed  to
unit, and the  science images were then super-flat corrected, aligned
to a single reference frame, and summed.  
 At the last step  we used only the best quality individual
 images  where stellar profile $FWHMs$ are $\la$ 0\farcs4.  
 This condition was satisfied  for 34  $H$  and  48    $K_s$  frames. 
 The resulting effective mean seeing values and integration times  
 were 0\farcs3,  0\farcs25  and 
7752,  10944 s  for the summed $H$ and $K_s$ images, respectively.

 Photometric standards (S273-E, S363-D, S705-D) from 
 \citet{1998AJ....116.2475P}  were observed  each night in both bands  for photometric 
 calibration. 
The airmass-corrected zero points provided by the NACO
pipeline (Table~\ref{t:log}) were used  
for photometry  of several bright stars  
 in individual science frames. Their magnitudes were then used  as secondary 
 standards for calibration of the summed images.
 The resulting magnitude zero-points  for the summed images are
 $C_H=24\fm14(5)$; $C_{K_s}=23\fm15(5)$.   
The errors   include  
uncertainties of  measurements and extinction coefficients, 
and marginal zero-point variations from night
to night. 

Astrometric referencing  of the summed   images in 
both 
 bands was performed with the IRAF {\it ccmap}  task. 
There are no suitable catalogue astrometric standards in the
field and  several   secondary astrometric standards  were
used instead. Their WCS coordinates were obtained
from  our VLT
optical images  of the same field referenced by 
\citet{2008A&A...492..805Z}.  Resulting {\sl rms} uncertainties of the  astrometric
fit  are $\le$ 0\farcs1 for RA and Dec. 

The resulting  $H$ and $K_s$  images   
are shown in Fig.\ref{fig2}.
Conservative $3\sigma$ detection limits  for a star-like object, 
$H_{3\sigma} = 23\fm4$  and $K_{3\sigma} = 22\fm3$, were 
derived  using  standard  $3\sigma$   background deviations 
within an optimal aperture of a radius of 8 pixels  
($\approx 0\farcs22$),  where the signal to noise ratio 
for a stellar-like object  reaches  
a maximum. The object  marked as "o1" is  one of the faintest 
objects reliably detected  in both bands with magnitudes    
$H^{o1}=23\fm95(20)$ and $K_s^{o1} = 22\fm75(20)$.
 Below we adopt 
 these magnitudes as  real detection limits.  
\subsection{The HST archival data}
\begin{figure}[t]
\setlength{\unitlength}{1mm}
\resizebox{7.cm}{!}{
\begin{picture}(70,63)(0,0)
\put (0,0) {\includegraphics[width=8.95cm, bb = 15 25 805 575, clip]{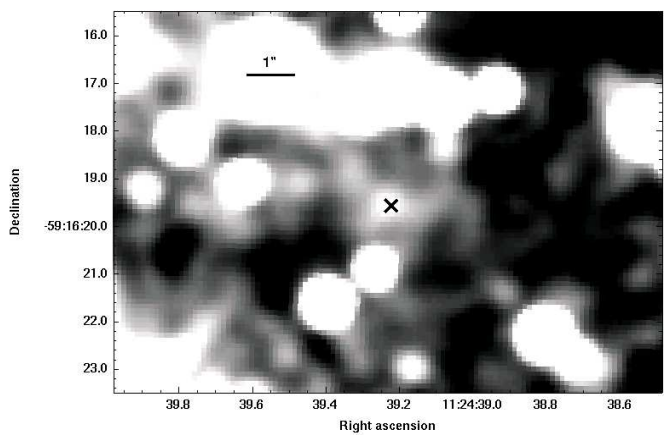}}
\end{picture}}
  \caption{Fragment of the HST/WFPC2/F675W
  archival image with the pulsar position indicated
  by "$\times$". 
The  image is smoothed with a 7-pix Gaussian kernel to clear the PWN, which is barely resolved,  while the pulsar is not. }
 \label{figX}
\end{figure}
To verify our optical fluxes of the pulsar+PWN system
obtained with the VLT \citep{2008A&A...492..805Z} we inspected the
HST archive and  found  
 two detailed observations  of
G292.0-1.8 in various filters with the  WFPC2 in 2007 and
2008.  Our system is detected  at $\sim$3$\sigma$
level only in  F675W filter images  
 obtained   in April 6 2008\footnote{GO 10916, PI R.
Fesen}  with the total integration time of 2400 s.  
The data are not published.  We combined the  pipe-line
reduced 
images and show the resulting image in Fig.\ref{figX}. 
\subsection{The AKARI archival data}
To update the mid-IR data on the pulsar+PWN system
emission reported early by  \citet{2009A&A...508..855Z}  based on the Spitzer
observations, we used the AKARI  archival data. 
The AKARI mid-IR imaging  of G292.0-1.8 was carried
out  in January 17  2007\footnote{Observation ID 1400749,
PI H. Kaneda} 
with the Infrared Camera (IRC) equipped with an {\it Si:As} detector
array with the pixel scale of 2\farcs51 and   $\approx$ 10\amin
$\times$10\amin~FOV. 
The  data with an analysis of the SNR emission 
have been published \citep{2009ApJ...706..441L}. 
For our analysis, we used only the images obtained with 
the filter centered on 15 $\mu m$ (IRC L15) where the
pulsar+PWN system is detected.  The total on-source integration 
time was 180~s. The basic calibration and data handling,
such as the dark-subtraction,
linearity fitting, distortion correction, flat-fielding,
and image combining, were performed  using the IRC Imaging Data Reduction 
Pipeline version 070908.

\section{Results}
\label{res}
%
%
\subsection{The pulsar+PWN near-IR counterpart}
An extended   source is clearly detected at the 
pulsar+PWN position in both $H$ and $K_s$ bands   
(Fig.\ref{fig2}). It is elongated
in the same SE-NW direction 
as the torus-like  part of
the PWN  in other spectral domains \citep[e.g.,][]{2008A&A...492..805Z}.  
We are thus confident,  that it is  the real near-IR counterpart  
of the pulsar+PWN system.  In the near-IR the PWN  torus appears to be more extended,  up to $\approx$3\farcs0, 
than in the optical, $\approx$2\farcs0, but is less extended than in the mid-IR,  $\approx$10\farcs5, and
X-rays,  $\lesssim$12\farcs \citep[cf.,][]{2009A&A...508..855Z}. 
 Therefore, in the near-IR, as in the optical, we see only its brightest inner 
part containing the pulsar.     

For the counterpart  photometry, 
we used an elliptical aperture with the center at the pulsar
X-ray position and  a rectangular box free of any sources to extract the
source emission and   backgrounds,  respectively (Fig.\ref{fig2}).  
The  resulting magnitudes are $H^{pwn+pulsar} = 21\fm30(10)$ and
$K_s^{pwn+pulsar}=20\fm45(10)$.    The real source  boundary 
is apparently different from the pure elliptical one by a various
way depending on the band. We account for this 
in the   magnitude errors,  which include not only statistical measurement and
calibration uncertainties  but also  flux variations  with 
a random shift of the ellipse center   within a circular area of
$\sim$ 5 pixels  radius  centered at the pulsar.

To complete the PWN photometry, 
the HST magnitude of the pulsar+PWN optical counterpart marked by "x" in  
Fig.\ref{figX}  was estimated as $m_{F647W} =24\fm2(3)$.  
This  marginal detection is  compatible with the R-band magnitude  
obtained by us from  deeper VLT observations \citep{2008A&A...492..805Z}.   
The magnitude of the system estimated from the AKARI 
 15 $\mu$m data is    $m_{15 \mu m}$=11\fm5(2).  This corresponds
 to the flux  $log~F_{\nu} [\mu Jy]= 2.6\pm0.08$. 
 It  may be  overestimated due to possible contribution of
 unresolved stars  located near the PWN boundary, but  
 can be  taken as a confident upper limit on 
 the system flux in this band.  
 The measured and  available observed counterpart  magnitudes, fluxes in physical units,  and de-reddened
 fluxes 
 for a most plausible interstellar extinction range
 of    1\fm86$\la$$A_V$$\la$2\fm10 \citep{2008A&A...492..805Z}  
 are  collected in Table~\ref{t:flux}. 
\begin{table}[t]
\caption{Observed magnitudes, fluxes and de-reddened fluxes  for the presumed optical/infrared PWN/pulsar counterpart of J1124-5916. }
\begin{tabular}{llll}
\hline\hline
$\lambda_{eff}$(band)            &  Mag.                 &  Log ($Flux_{obs}$)         & Log ($Flux_{A_{V}}$)         \\
    ($\mu$ m)                          &  observed           &     ($\mu$Jy)                     & ($\mu$Jy)                            \\  \hline \hline                 
    0.55(V)                               & 24.29(13)           & -0.16(5)                          & 0.66($^{+7}_{-13}$)          \\
    0.66(R)                               & 24.12(13)           &-0.17(5)                           & 0.51($^{+7}_{-12}$)          \\
    0.67(F675W$^\dagger$)      & 24.2(3)              &  -0.2(1)                            & 0.57($^{+16}_{-17}$)         \\
    0.77(I)                                &23.12(13)           & 0.13(5)                            & 0.66($^{+7}_{-12}$)           \\
    1.65({\bf H})                       & 21.30(10)          &  0.52(3)                           &   0.66($^{+6}_{-10}$)       \\
    2.16({\bf K$_s$})                 &20.45(10)           &  0.66(3)                          &    0.75($^{+6}_{-10}$)           \\
    4.5                                      &15.9(4)              & 1.90(21)                          &   1.95( $^{+21}_{-22}$)         \\
    8.0                                      &14.2(3)              &  2.13(16)                         &   2.18($^{+16}_{-17}$)        \\
    15                                       &11.5(2)                  & 2.60(8)                           &   2.60(8)               \\
    24                                       &$\ge10.2$           & $\le 2.8$                        &     $\le2.85$                          \\
    70                                       &$\ge6.8$             & $\le 3.2$                        &     $\le3.2$                    \\
 \hline   
 \end{tabular} \\
 
{\footnotesize $A_V$ = 1.86 --- 2.10 \\
 $^\dagger${Our estimate using the archival HST/WFPC2/F675W data}
 } 
 \label{t:flux}
\end{table}

\begin{figure*}[t]
\setlength{\unitlength}{1mm}
\resizebox{12.cm}{!}{
\begin{picture}(120,65)(0,0)
\put (0,0) {\includegraphics[width=9.4cm,  angle=0, clip]{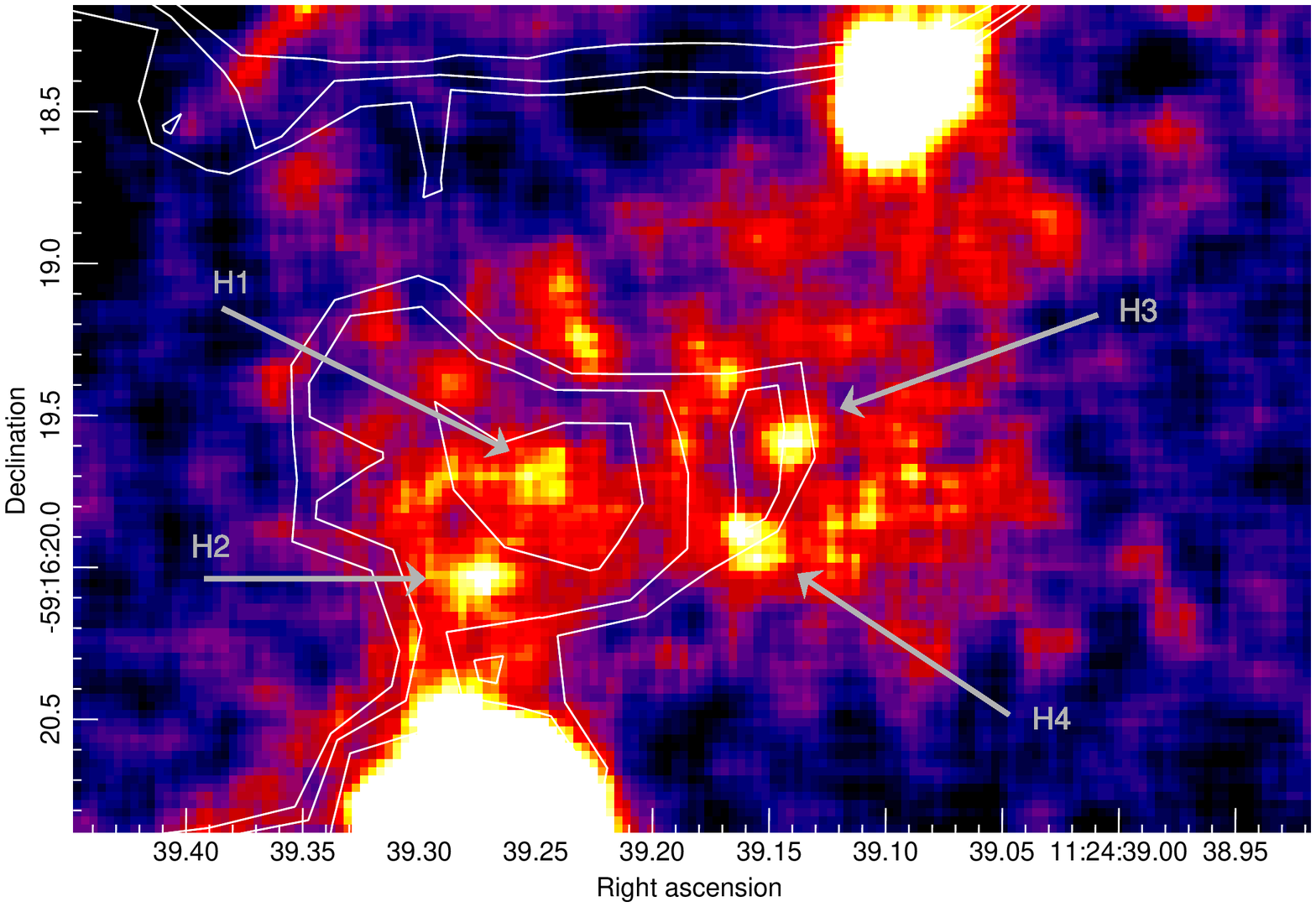}}
\put (90,0) {\includegraphics[width=9.4 cm, angle=0, clip]{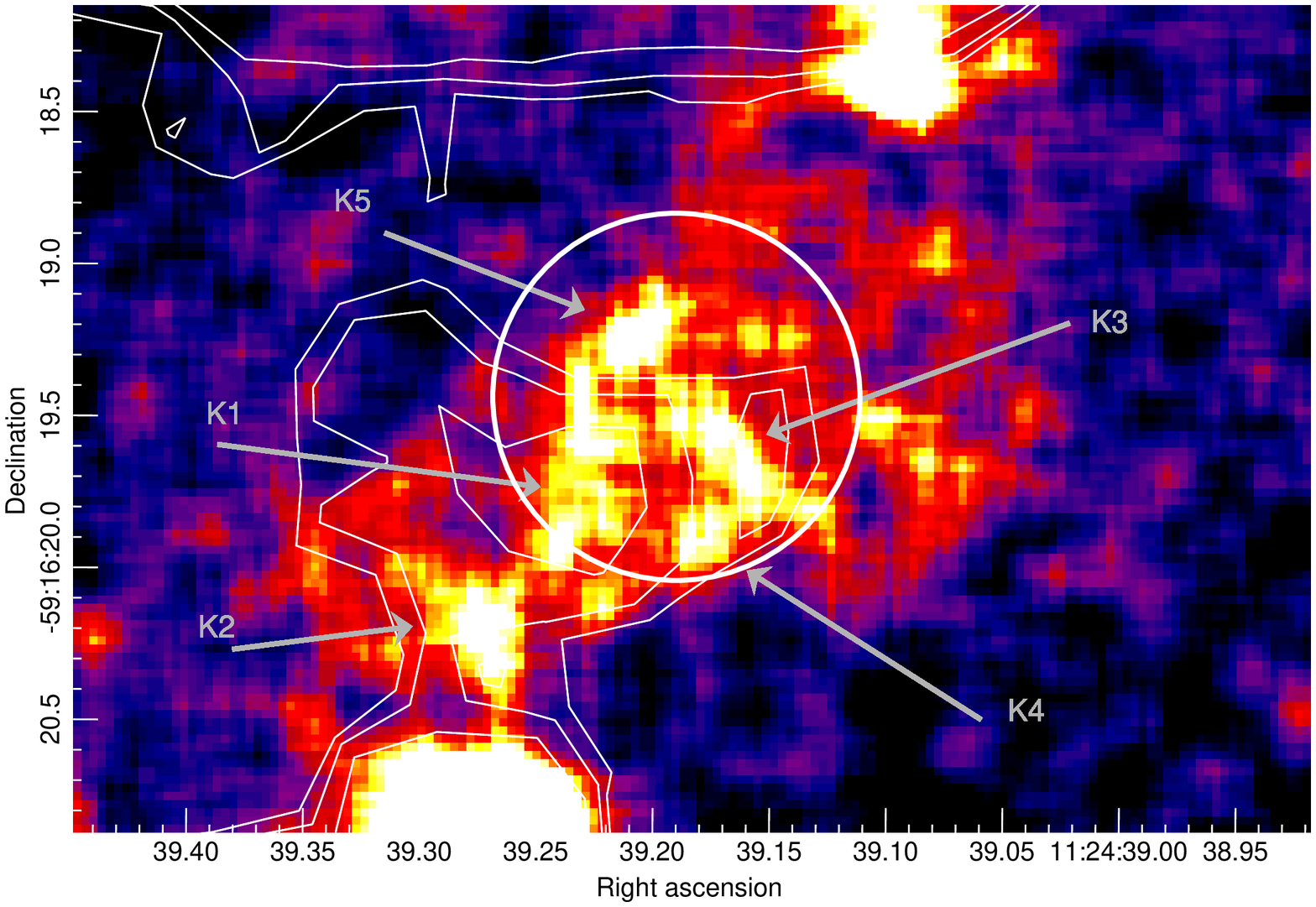}}
\end{picture}}
  \caption{$H$ ({\sl left}) and $K_s$ ({\sl
  right})  image fragments zoomed in on  the
  X-ray pulsar position whose 1$\sigma$ error is shown by the
  circle. The images are   
  smoothed  with  a Gaussian $3\times3$ pixel
  kernel, and  their horizontal  sizes are   
  $\sim$4\farcs5. 
  As in Fig.~\ref{fig2}, the contours are  from the 
  $V$ band image, they show the  structure of
  the PWN in the optical.
  Several relatively bright knots nearly the pulsar
  position  are marked by H1-H4 and K1-K5.
   }
 \label{fig3}
\end{figure*}
\subsection{PWN knots and searching for the pulsar counterpart}

 As seen from  zoomed in fragments of the near-IR images   (Fig.~\ref{fig3}),  the PWN brightness spatial
distribution is not uniform.  Some
relatively bright "knots", marked as H1---H4  and K1---K5, are visible
within  the 1$\sigma$ error-circle  of the  pulsar X-ray position,   where  the maximum  
of the   brightness in the optical and
X-rays is located. The knots
are  resolved neither in the optical  nor
in X-rays
due to significantly lower spatial resolution or  
a shallow exposure, as in the HST case. 
They cannot be due to  the detector noise, 
since such structures are not visible in other image parts.
Not all of them are  surely cross-identified in both bands, while  
 the  pairs H1/K1, H2/K2, and H4/K4  likely  have the same origins. 
The knot H1/K1 is located just at the PWN optical brightness maximum   
and  could be a candidate to the pulsar counterpart. 
Others  are likely to be compact synchrotron structures  formed in the PWN torus 
as is in the Crab  case. 
Alternatively, the knots might be a SN ejecta that has been shocked by the PWN.  For instance, 
\citet{2012A&A...542A..12Z} reported on FeII emission from 
G21.9-0.9 in the 1.64 $\mu$m narrow band overlapping with the $H$ band. 
It forms a shell-like structure around the  G21.9-0.9 PWN X-ray boundary. 
Similar narrow band studies of the G292.0+1.8 field are necessary to  
establish whether the knots visible inside its PWN torus  are related to the shocked  
ejecta or not.  
The knots could also be faint unrelated background sources projected  on the PWN.   
However, this is very unlikely due to a much smaller surface density of such 
sources in other parts of the field. 

The local background in the PWN area is  higher  than
that around faintest isolated objects outside it, for
instance, around source o1  in
Fig.~\ref{fig2}.  This precludes us to
confidently  
 measure  the knot magnitudes
 relative to the local background with an 
 accuracy of better than 0\fm3.      
A conservative  estimate yields  
that  their fluxes are about or a few less 
of the  source    o1 flux,   which is
firmly  detected  at a 5$\sigma$ 
significance in both bands (Sect. 2.1).  
Based on that, the knot and 
pulsar brightness upper limits are 
 $H^{psr} \ga 23\fm9$ and $K_s^{psr}
 \ga 22\fm7$.     
The  contribution of the cross-identified knots 
 H1/K1, H2/K2, and H4/K4 to the derived total pulsar+PWN-torus fluxes  
is estimated to be  $\lesssim$30\%.  Since the SN ejecta origin 
of the knots is questionable and they do not dominate  the entire pulsar+PWN flux,  
bellow we focus only on the analysis of the spatially integrated fluxes derived above.

\begin{figure}[t]
\setlength{\unitlength}{1mm}
\begin{picture}(120,212)(0,0)
\put (9.75,212) {\includegraphics[width=4.1cm, angle=-90, clip]{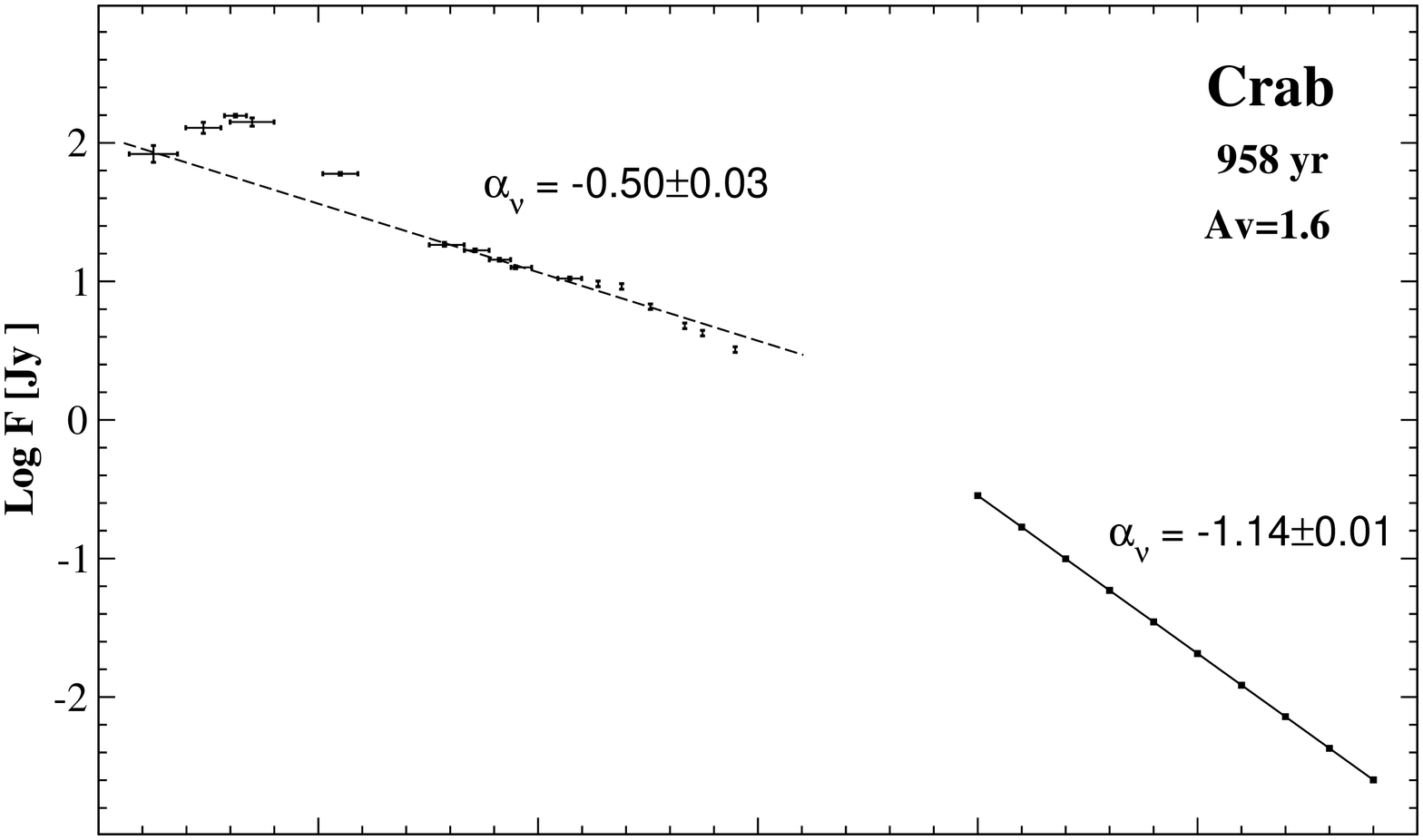}}
\put (9.5,170) {\includegraphics[width=4.1 cm, angle=-90,clip]{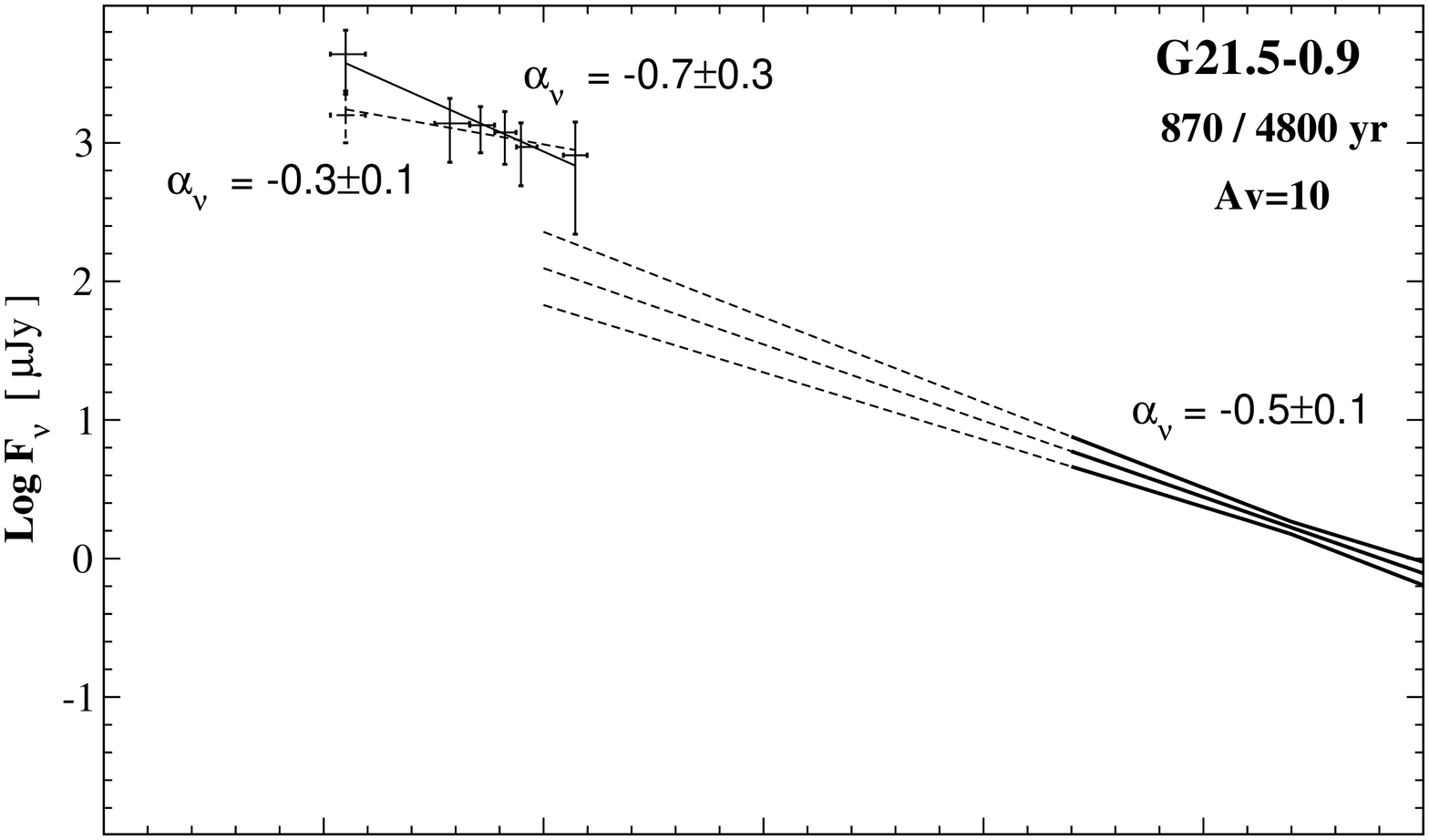}}
\put (9.73,128) {\includegraphics[width=4.1 cm,angle=-90,clip]{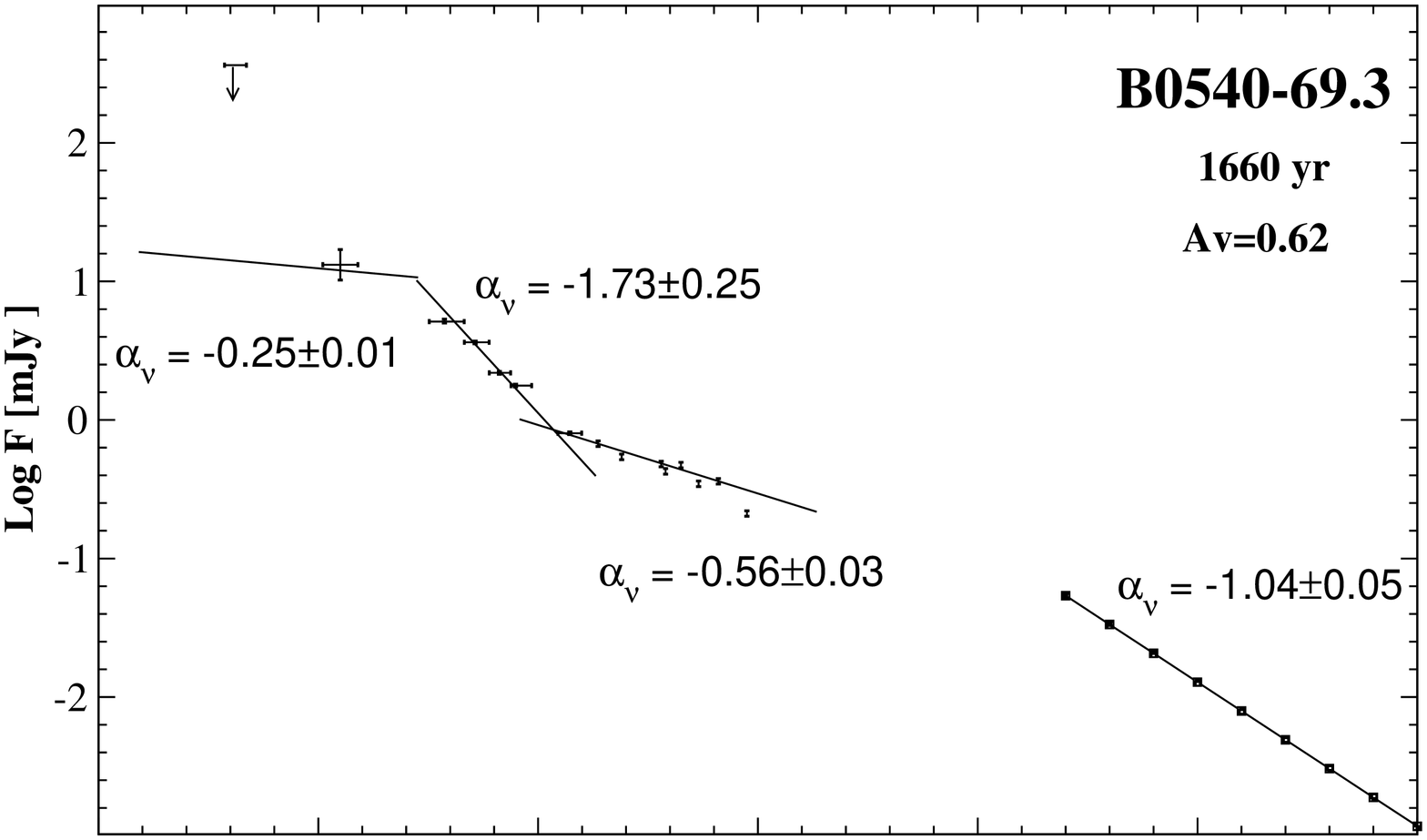}}
\put (9.5,86) {\includegraphics[width=4.1 cm,angle=-90,clip]{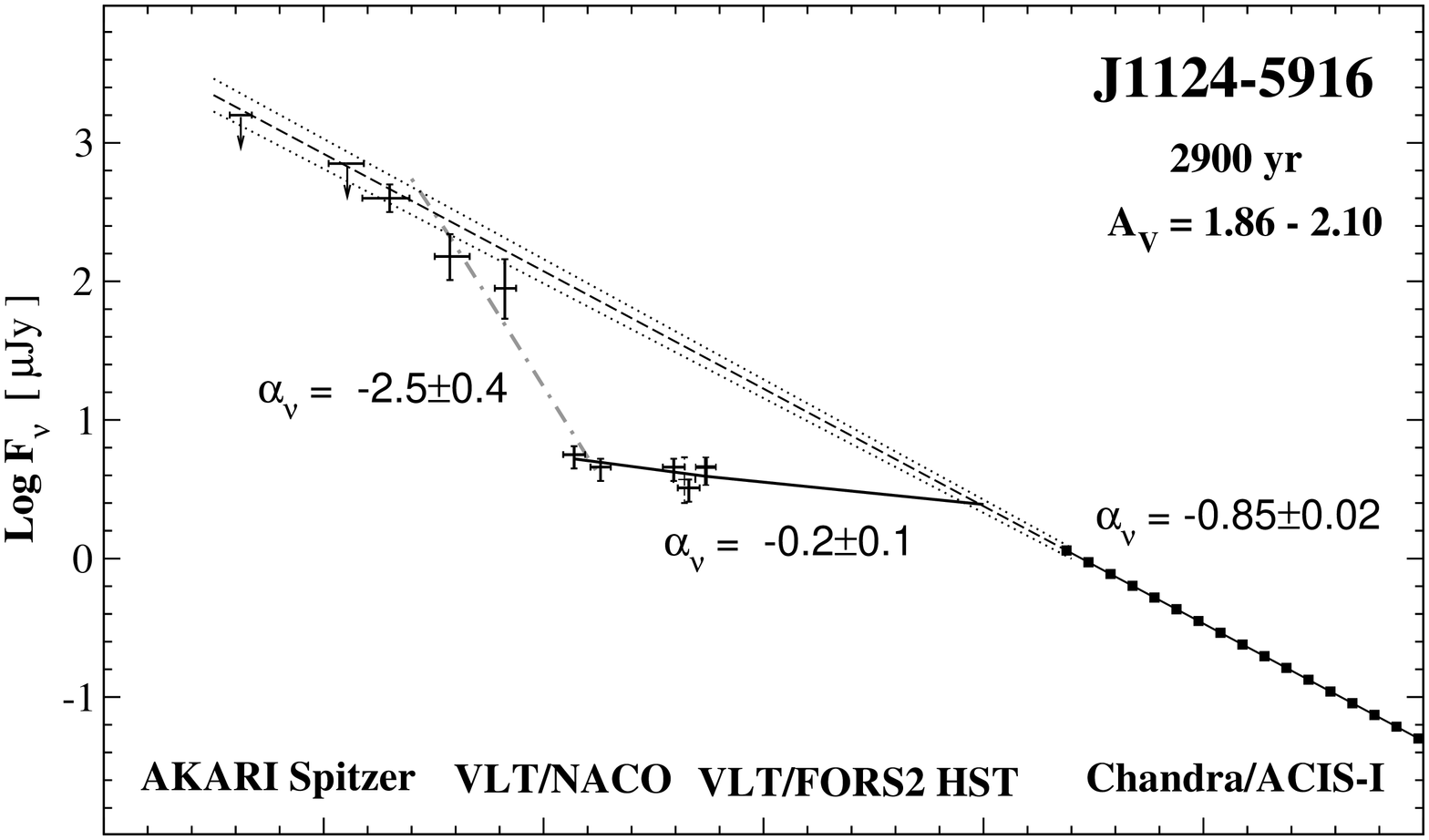}}
\put (9.5,44) {\includegraphics[width=4.54 cm,angle=-90,clip]{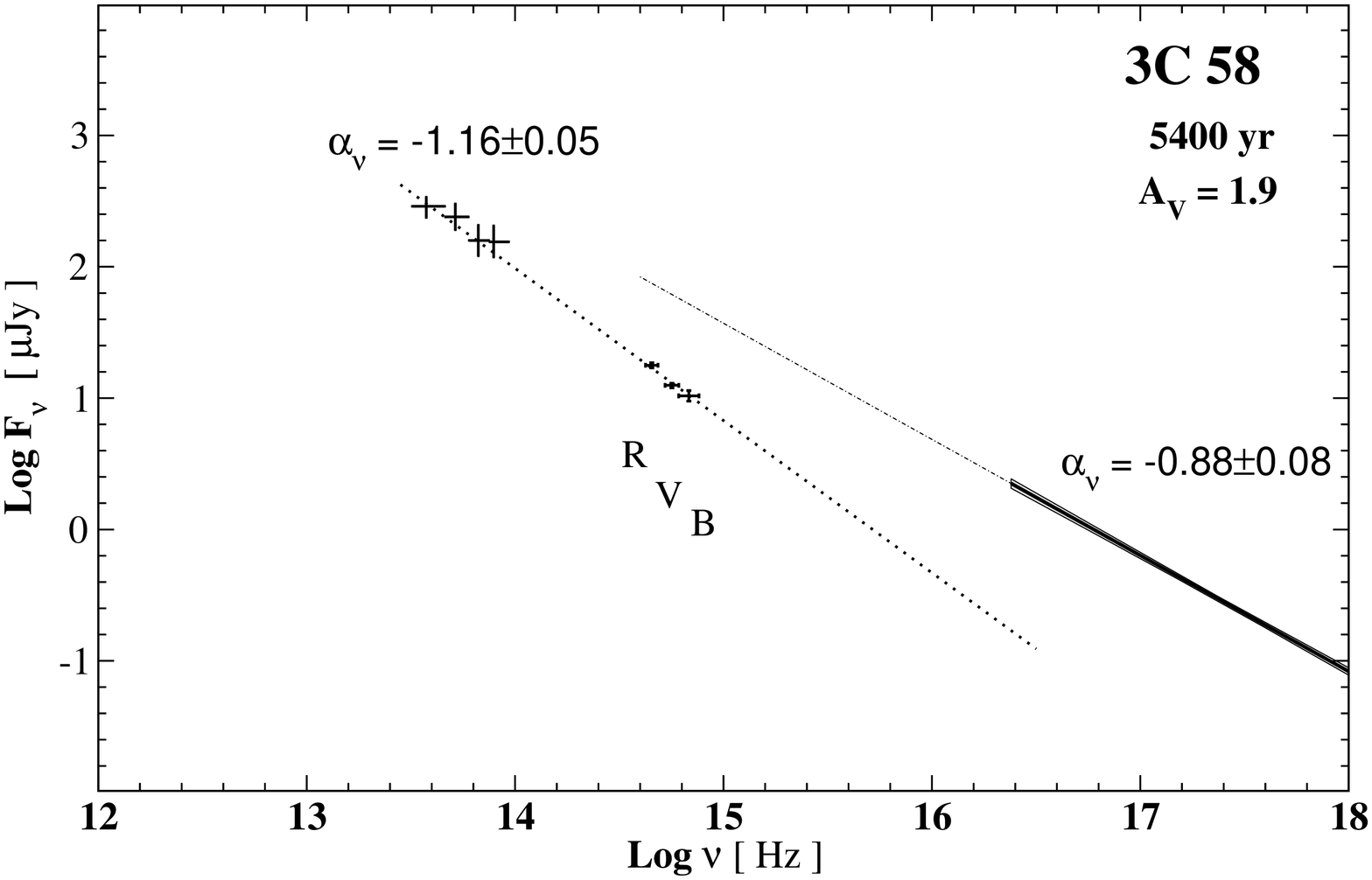}}
\end{picture} 
  \caption{Multi-wavelength spectra
   of torus-like
  PWNe obtained with different instruments and 
  ordered by PWN age from top to bottom. 
    }
  \label{fig4}
\end{figure}
\section{Discussion}
\label{dis}
 VLT NACO  observations allowed us to detect  the near-IR
counterpart  of the compact torus-like structure of the J1124-5916 PWN in the $H$ and $K_s$  bands 
and measure its spatially integrated fluxes. This 
makes J1124-5916 the fifth, after the Crab \citep{1979PASP...91..436G}, G21.5-09 \citep{2012A&A...542A..12Z}, 
B540-69 \citep{2012A&A...544A.100M}, and 3C 58
(Zyuzin et al., in preparation),  among other
PWNe ever identified in the near-IR.   
   
 A high spatial resolution of $\la$ 0\farcs3 provided by the  NACO,
allowed us to  resolve a fine structure of
the PWN whose spatial brightness distribution is  nonuniform. 
It contains several  relatively bright knots  located
within or near the pulsar 1$\sigma$ X-ray position error circle. Three of them are
cross-identified in the $H$ and $K_s$ bands,  and one of 
threes, H1/K1, is possibly associated with the pulsar.      
A high PWN background precludes a  confident knot flux
extraction,    
and we derive only flux  upper limits,  
 showing  
that the pulsar contribution to the total 
pulsar+PWN flux in the near-IR   is $\la$ 10\%. In
addition, the AKARI and HST archival data allowed us to constrain the
PWN mid-IR flux at 15 $\mu$m and confirm its optical
flux. Below we discuss 
possible  implications of 
the results.  

\subsection{Multiwavelength spectrum of  the torus-like PWN}
In Fig.~\ref{fig4} the  spectrum 
of the J1124-5916  torus PWN compiled from the data obtained here and 
published early by \citet{2009A&A...508..855Z}
  is  compared with available 
  spectra\footnote{The data are taken:
  for the Crab from  \citet{2004MNRAS.355.1315G}, 
  \citet{2006AJ....132.1610T},
  \citet{1979PASP...91..436G}, \citet{1993A&A...270..370V}, 
  and \citet{2005SPIE.5898...22K}; for  
G21.5-0.9 from \citet{2012A&A...542A..12Z};
for B0540-69.3 from \citet{1993ApJ...411..756M}, 
\citet{2008ApJ...687.1054W}, \citet{2012A&A...544A.100M}, and 
\citet{2005AdSpR..35.1106S};
for 3C 58 from \citet{2008A&A...486..273S},
\citet{2008MNRAS.390..235S}, and
\citet{2008ApJ...676L..33S}.} for other four torus-like PWNe
 mentioned above. 
   All of them, except for a highly absorbed G21.5-09, are also identified 
 in the optical. The spectra are non-thermal, as expected for the synchrotron origin of the PWN emission, 
 and can be described by power laws with
 different spectral indexes  $\alpha_\nu$ (defined as $F_\nu \propto
 \nu^{\alpha_\nu}$) depending on the spectral domain. 
  All of them show a strong   flux  increase towards   low frequency range.  
 For the youngest   Crab, and  possible for
 G21.5-09\footnote{We indicate in the plot the SNR
 \citep[870 yr,][]{2008MNRAS.386.1411B} 
 and spindown \citep[4800 yr,][]{2006ApJ...637..456C}
 ages for this PWN.},  the increase  appears to be monotonic 
 with a gradual  spectral slope  decrease
 towards the IR\footnote{A  mid-IR bump over the
 power law fit  in  the Crab 
  spectrum in the mid-IR at  $\sim$50  $\mu$m is due
  to  thermal emission from a warm dust created 
  by the SN  \citep{2006AJ....132.1610T,
  2011ApJ...734...54A, 2012ApJ...753...72T}.}.    While for the older J1124-5916 
   we see at least two strong  spectral breaks:  one is between  the optical and X-rays and another 
 is between the near-IR and mid-IR.   The third
 possible break   exists above 15   $\mu$m with
 slope flattening towards lower frequencies.  
 A sharp, likely a double knee break between the
 optical and X-rays is visible in  the spectrum 
 of a similar age 3C 58 PWN   \citep{2008A&A...486..273S}\footnote{The optical-IR
 part of the 3C 58 torus spectrum is confirmed by near-UV and near-IR
 observations (Zyuzin et al., in preparation).}. 
The spectrum of  B0540-69, which is   slightly older  than the Crab,  becomes   apparently steeper 
in the mid-IR, as compared to   the optical-near-IR,
suggesting a break at a few microns
\citep{2012A&A...544A.100M}, and then flattens  again
at longer wavelengths \citep{2008ApJ...687.1054W}. A double
knee break between the optical and X-rays is also
not excluded \citep{2005AdSpR..35.1106S}. 
At the same time, the IR part of the  G21.5-09 
spectrum is likely to be 
almost flat\footnote{ We remeasured the Spitzer
fluxes for this PWN and found a good agreement with 
\citet{2012A&A...542A..12Z} for all bands  except for 24 $\mu$m, 
where our flux of $(1.58
\pm 0.81)$ mJy,   is  by a factor
of 3 lower than  their one and consistent with more 
flat spectral index  derived using lower wavelength bands.
The reasons for the discrepancy are unclear, and 
in Fig.~\ref{fig4} we show both measurements for this
band and respective fits.}, 
implying at least one break between the IR and 
X-rays \citep{2012A&A...542A..12Z}.

 \begin{table*}[t]
\caption{$H$ and $K_s$ magnitudes, 
near-IR luminosities $L_{IR}$, and radiation
efficiencies  $\eta_{IR}$ of radio pulsars observed 
in the near-IR.}
\begin{tabular}{lllllllll}
\hline\hline
Pulsar name	      & $\log \tau$ & $\log \dot{E}$
&  distance                               &  $A_V$  
     & H, K$_s$ mag      & $\log L_{IR}$   & $\log
     \eta_{IR}$   & Ref.  \\ \hline \hline
    Crab	           &   3.1     &	38.65	  &  1.73(28)                              & 1.62            & 	14.27(5), 13.77(5)	 & 32.6(1), 32.9(1)  	& -6.1(1), \hspace{1mm} -5.8(1) &	(1)	\\
    PSR 0540-69  &  3.2     &  38.17        &   5$\times$10$^4$               &   0.62                 &     19.33(10), 18.55(10)  & 33.2(1), 33.3(1)       &        -5.0(1), -4.9(1)   &      (2)     \\
    PSR B1509-58  &   3.2     &  37.25     &  5.2(1.4)                              & 4.8              & 20.6(2),\hspace{2.5mm}19.4(1)&31.1(2), 31.6(1) & -6.2(2), \hspace{1mm} -5.7(1) & (3) \\ 
    PSR 1124-5916 &  3.5     &  37.08     &   $\sim6.0$                           & 1.98(12)      & $\ge 23.9$,\hspace{2mm} $\ge 22.7$ &  $\le29.8$, $\le 29.9$            & $\le$-7.3,\hspace{3mm} $\le$ -7.2 	& tw	\\
   Vela                &	  4.1     &  36.84	&0.293$^{+0.019}_{-0.017}$   & 0.20	            &  22.04(16), 21.3(4)                 & 27.7(1), 28.1(2)             &	-9.1(1),\hspace{1mm}  -8.7(2)	& (4)	\\
   PSR B0656+14$^*$ &   5.0    &  34.58	&0.288$^{0.033}_{-0.027}$     & 0.09(6)        & 23.22(8),\hspace{1.6mm} 22.63(13)    &	27.3(1), 27.6(1) &	-7.3(1),\hspace{1mm}  -7.0(1)	& (5)	\\
   Geminga$^*$          &   5.5    &  34.51	& 0.157$^{+0.059}_{-0.034}$  & 0.12(9)       &  24.30(10), 23.4(4)                &   26.3(1), 26.7(2)            & -8.2(1),\hspace{1mm}    -7.7(2)	& (5)	\\
 \hline   
 \end{tabular} \\
 \label{t:IR}
 ($^*$) based on the HST F110W and F160W band
 observations; \\
 (1) \cite{2003A&A...406..639S}; (2) \cite{2012A&A...544A.100M}; (3) \cite{2006ApJ...644.1056K}; (4) \cite{2003A&A...406..645S}; (5)   \cite{2001A&A...370.1004K}
  \end{table*}

We can conclude that the presence  of several
prominent spectral breaks  in the IR-optical-X-ray range  is  a common
feature for    torus-like PWNe.  This cannot be excluded even for the Crab, 
where, in addition to the significant X-ray-optical  
slope change, some small   breaks may exist between the optical and near-IR and/or 
between the near-IR and mid-IR \citep{2011ApJ...734...54A}. 
In   Fig.~\ref{fig4} the  breaks likely become 
stronger with PWN age.  
However, this apparent spectral evolution trend has to be
considered with a caution, since the    
spectrum of the Crab is dominated by 
the small torus  region only in X-rays, while at  low
frequencies it includes  the emission from the whole
plerion resulting to the smoothness of the spectrum.   
The same problem is not solved yet for the Crab "twin", B0540-69, 
whose   PWN torus is not yet  firmly resolved and its position and orientation 
are still debated \citep{2007ApJ...667L..77D,2011MNRAS.413..611L}.  

\begin{figure}[t]
\setlength{\unitlength}{1mm}
\begin{picture}(120,65)(0,0)
\put (0,0) {\includegraphics[width=9cm, angle=0, clip]{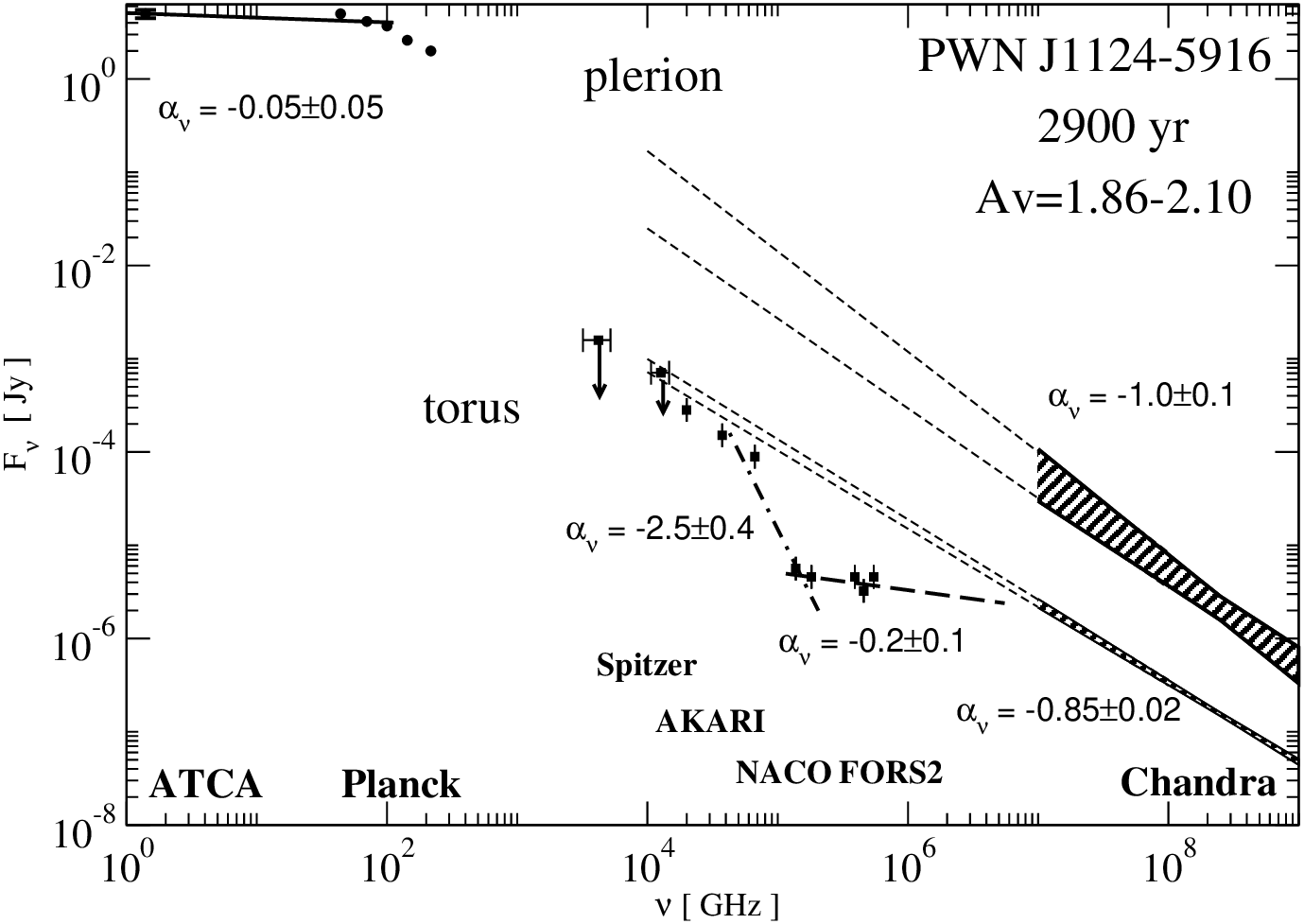}}
\end{picture} 
  \caption{ Comparison of the spatially integrated multiwavelength spectra of the J1124-5916 PWN  
    for the compact PWN torus (lower curves)  and for the much more extended plerion (upper curves) obtained with 
    different instruments, as notified in the plot. The radio and submillimetre data for the plerion are taken from  
    \citet{2003ApJ...594..326G} and \citet{2011A&A...536A...7P}, respectively. 1$\sigma$ X-ray uncertainties for each of 
    the spectrum  are 
    indicated by hatched regions and they are extended towards low frequency range.    
    The torus is not detected in the radio, while the plerion is not identified  
    in the IR.
    }
  \label{fig5}
\end{figure}
   
 It is evident that the compact torus-like PWN works as an energy injector for the large scale plerion. 
As mentioned by
\citet{2003ApJ...594..326G}, the spectrum
of the  J1124-5916 large scale plerion has  
a single spectral break between the radio
and X-rays with $\Delta \alpha_{\nu} \sim 1$ at a relatively low
frequency range of $\la 800$ GHz.
The Planck early release  compact source catalog\footnote{ astro-ph arXiv:1101.2041}   allows us to
constrain more precisely the break  position near 70 GHz, where the  plerion flux of   $(4.1 \pm
0.2)$  mJy becomes significantly lower than that of $(5.3 \pm 0.1)$  mJy at  5 GHz measured early,
and continues to drop  rapidly   with the frequency  (e.g., $(2.1 \pm 0.1)$ mJy  at 217 GHz; { Fig. \ref{fig5}}).
{ In Fig.~\ref{fig5} 
the X-ray spectrum of the  G292.0+1.8 plerion  was obtained using the data from 
the Chandra archive\footnote{ID 6677, PI S. Park}. 
We reprocessed ''level 1'' data with CIAO chandra\_repro script and used the spectral interval  of 5 -- 10 keV 
to exclude the G292.0+1.8 thermal component strongly dominating  the nonthermal plerion spectrum 
at smaller energies. 
The spectrum was extracted using a source aperture of 1\famin1 centered  at the pulsar, background was taken 
outside the SNR,  
and  fitted by a single absorbed power law with a resulting slope  $\alpha_\nu =-1.0(1)$.} 
The low frequency spectral break of the plerion cannot be simply explained  by synchrotron losses 
of a single particle population \citep{2003ApJ...594..326G}, suggesting,  as is in the Crab   
\citep{1984ApJ...283..710K}  { and 3C 58 \citep{2008ApJ...676L..33S} cases}, 
distinct populations for the radio and X-ray  emission of the plerion. 
The observed multiple spectral breaks in
the torus emission serve as the direct evidence of 
a complicated spectrum of the injected
particles, which cannot be described by a simple model
of an unbroken powerlaw. This fact can help
us to better  understand the nature of the low-frequency
break of the large scale plerion for   
J1124-5916 and similar breaks for other
PWNe.  To do that more detail observations of PWNe and 
 modeling of these systems are necessary.

\begin{figure}[t]
\setlength{\unitlength}{1mm}
\resizebox{12.cm}{!}{
\begin{picture}(100,55)(0,0)
\put (0,0) {\includegraphics[width=7.0cm, clip]{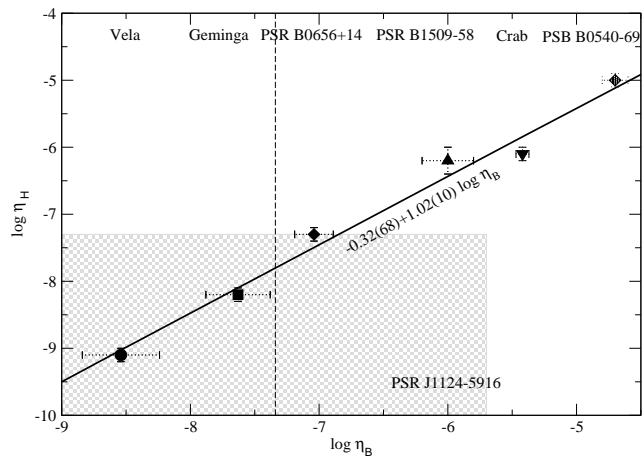}}

\end{picture}}
  \caption{  
  Correlation between the optical ($B$ band) and
  near-IR ($H$ band) efficiencies of radio pulsars. 
  The optical data are taken from \citet{2006AdSpR..37.1979Z}.   
  }
 \label{fig6}
\end{figure}
%
\subsection{Pulsar flux upper limits and
efficiencies}
In Table. \ref{t:IR} we summarize  observed  $H$,
$K_s$ magnitudes, respective luminosities and radiation efficiencies 
$\eta = L_{IR}/\dot{E}$  of  rotation powered pulsars detected in
the near-IR and optical bands. 
 The $H$   {\sl vs.} $B$ band efficiency plot (Fig.~\ref{fig6}) shows a strong correlation between the
 optical and near-IR  efficiencies. This  confirms a common origin of radiation  in the
 near-IR and   the optical resulting also
 from rather smooth
 pulsar spectral energy distributions in these ranges.  The 
 $\eta_{H} \propto \eta_{B}$ empirical relationship derived
 from the plot (thick solid  line) allows  
 one to predict  IR magnitudes for all pulsars which have been 
  detected in the optical but not yet in the near-IR  and  {\sl vice versa}.   
  
  Based on the observed  PSR J1124-5916 flux upper
  limits in the near-IR obtained here and in the
  optical provided by  \citet{2008A&A...492..805Z}, 
   we  can constrain the pulsar  position region  on the efficiency plain (shaded region in  Fig.~\ref{fig6}) 
    and conclude   that  its  efficiencies  are  at least  lower  than those of younger 
   pulsars.  The  $\eta_{H}\propto \eta_{B}$
   relationship  and the $H$ band limit suggest 
   that the pulsar cannot be more efficient than
   the middle-aged  PSR B0656+14, and allows us to
   better constrain its  $\eta_{B}$  (vertical
   dashed line) than it was possible from
   our optical observations.  We predict
   that its  de-reddened  $B\geq$26\fm8. This   
  means also that the pulsar contribution to the
  total pulsar+PWN optical flux is $\le 10$\%,  as
  is in the near-IR.
 \section{Conclusions}
\label{conc} 
Using the NACO at the VLT we performed the deepest
up to date near-IR observations of the young pulsar J1124-5916 field.
In the $H$ and $K_s$ bands we detected a faint $H = 21.30(10)$, $K_s = 20.45(10)$ extended 
elliptical object, whose center position is consistent with the X-ray position of the pulsar. 
The morphology of the object and the orientation of its major axis are in a good agreement 
with those for the J1124-5916 torus-like PWN
obtained in the optical and X-rays. This allows us to conclude that we
detected the near-IR counterpart of the  PWN.
The compiled IR-optical-X-ray power-law spectrum of
the torus-like PWN shows several spectral breaks implying a
multiple population of relativistic particles
responsible  for the emission in different spectral
domains. This may help to explain a single low
frequency break in the radio-X-ray spectrum of the large scale 
plerion of J1124-5916, which is impossible to describe by a single
particle population.  The presence of several
spectral breaks appears to be a common feature for
the spectra of all torus-like PWNe. 

We found several faint  knot-like objects   nearly
the X-ray position of the pulsar.  
The PWN background is high in this area, which has not 
allowed us  to conclude  confidently about their
origin. We derived  the upper limits for the pulsar 
near-IR fluxes and estimated  its contribution to the
total   pulsar+PWN  flux in this range at a level of $\leq10$\%. 
Comparing to other pulsars observed in the near-IR
and optical we conclude, that the expected pulsar
contribution in the optical  should  not also exceed this
level.  Deeper  high-spatial resolution optical
and near-IR observations  are necessary to resolve
the pulsar from its PWN. 
\begin{acknowledgements} 
      The work was partially  supported by CONACYT
      151858  projects and by the Russian Foundation for Basic 
Research (grants 11-02-00253 and 13-02-12017-ofi-m), RF Presidential Program (Grant NSh 4035.2012.2). 
REM acknowledges support by the BASAL Centro de Astrofisica y Tecnologias Afines (CATA) PFB--06/2007. 
 \end{acknowledgements}

\bibliographystyle{aa} 
\bibliography{szharikov.bib}
\end{document}